\documentclass[review,12pt]{elsarticle}
\usepackage{lineno,hyperref}
\modulolinenumbers[5]
\usepackage{amsmath}
\usepackage{amsfonts}
\usepackage{amssymb}
\usepackage{lmodern}
\usepackage{xfrac}
\usepackage{graphicx}
\graphicspath{ {./figures/} }
\usepackage{epsfig}
\usepackage{epstopdf}
\usepackage{textcomp}
\usepackage{multirow}
\usepackage{setspace}
\usepackage{fullpage}
\usepackage{float}
\usepackage{caption}
\usepackage{subcaption}
\usepackage{siunitx}
\usepackage{xcolor}

\biboptions{sort&compress}

\begin{document}

\begin{frontmatter}
\title{Simulation and analysis of $\gamma$-Ni cellular growth during laser powder deposition of Ni-based superalloys}

\author[mymainaddress]{Supriyo Ghosh\corref{mycorrespondingauthor}}
\ead{supriyo.ghosh@nist.gov}
\author[mymainaddress,mysecondaryaddress]{Nana Ofori-Opoku}
\author[mymainaddress]{Jonathan E. Guyer\corref{mycorrespondingauthor}}
\ead{jonathan.guyer@nist.gov}
\address[mymainaddress]{Materials Science and Engineering Division, National Institute of Standards and Technology, Gaithersburg, MD 20899, USA}
\cortext[mycorrespondingauthor]{Corresponding author.}


%
\address[mysecondaryaddress]{Center for Hierarchical Materials Design, Northwestern University, Evanston, IL 60208, USA}


\begin{abstract}
Cellular or dendritic microstructures that result as a function of additive manufacturing solidification conditions in a Ni-based melt pool are simulated in the present work using three-dimensional phase-field simulations. A macroscopic thermal model is used to obtain the temperature gradient $G$ and the solidification velocity $V$ which are provided as inputs to the phase-field model. We extract the cell spacings, cell core compositions, and cell tip as well as mushy zone temperatures from the simulated microstructures as a function of $V$. Cell spacings are compared with different scaling laws that correlate to the solidification conditions and approximated by $G^{-m}V^{-n}$. Cell core compositions are compared with the analytical solutions of a dendrite growth theory and found to be in good agreement. Through analysis of the mushy zone, we extract a characteristic bridging plane, where the primary $\gamma$ phase coalesces across the intercellular liquid channels at a $\gamma$ fraction between 0.6 and 0.7. The temperature and the $\gamma$ fraction in this plane are found to decrease with increasing $V$. The simulated microstructural features are significant as they can be used as inputs for the simulation of subsequent heat treatment processes.
\end{abstract}


\end{frontmatter}

\newpage
\section{Introduction}\label{sec_intro}
Ni-based superalloys possess excellent mechanical properties at elevated temperatures, which make them suitable for gas-turbine and jet-engine components \cite{Reed2008}. The laser powder bed fusion (L-PBF) additive manufacturing (AM) process is used to produce near net-shaped metallic parts from the alloy powder, in a layer-by-layer fashion with multi-pass laser melting, solidification and solid-state phase transformations, in a shorter manufacturing time than traditional casting, metal forming, and machining processes, with almost no waste and at a reasonable cost~\cite{Murr2012,Attallah2016}. Solidification in this process controls the size, shape and distribution of the $\gamma$ grains, the growth morphology, the elemental segregation and precipitation, the solid-state transformations and ultimately the properties and performance of the product. Understanding and predicting the melt pool solidification behavior is therefore important.

Additive manufacturing is a relatively new research field. Some of the outstanding issues in this process are related to the cell/dendrite spacing, microsegregation and residual stress \cite{Attallah2016} in the as-deposited parts. Several experimental investigations \cite{Formenti2005,Murr2012,Harrison2015,gaumann2001,Whitesell2000,Amato2012} and numerical simulations \cite{Lee2014,Liang2016,nie2014,Wang2003,Lee2010} have linked the AM process parameters to the melt pool solidification conditions -- temperature gradient $G$ and solidification velocity $V$ -- on a microscopic scale. The growth morphologies that result as a function of the solidification conditions and intercellular segregation determine the yield strength, ultimate tensile strength and fatigue strength of the material~\cite{Rappazbook}. The solidification conditions considered in previous studies have been relatively small-valued. For example the cooling rates, $\dot{T} = GV$, were reported to be no higher than 10$^4$ K s$^{-1}$, whereas the present work treats larger values ($\dot{T}$ = 10$^6$ K s$^{-1}$) of the melt pool solidification conditions. Therefore, the characteristic cell spacings, concentrations and temperatures in the parameter-microstructure map are expected to be different from those reported in the existing literature. 

Cell spacing depends on melt pool solidification conditions $G$ and $V$~\cite{kurzbook,Rappazbook,Whitesell2000} and is often estimated as $\lambda_c = AG^{-m}V^{-n}$, where $m$ and $n$ are the model-dependent exponents and the coefficient $A$ depends on alloy properties. During solidification, solute is partitioned between the solid and liquid phases, and in ideal conditions reaches the corresponding equilibrium values of the phase diagram. Such solute partitioning however is incomplete in the AM solidification regime and therefore the solute concentration field predicted by the phase diagram no longer applies~\cite{Liang2016,supriyo2017}. In order to reflect this departure from equilibrium, solidification parameters in the theories are phenomenologically modified from equilibrium to velocity-dependent values~\cite{Kurz1994,Liang2016}. The above theories are used as a reference to compare with our spacing and microsegregation simulation data. 

The mushy zone in cellular microstructures is a two-phase solid and liquid coexistence region between the
fully solid and the fully liquid states where the majority of the solidification defects form \cite{Rappaz1999,Wang2004}. These defects
arise due to the random growth of the solid cells towards one another which finally coalesce and thereby restrict the feeding of the liquid to accommodate shrinkage during late solidification stages. As the solid fraction in the mushy zone increases, the liquid is not able to flow freely and compensate for shrinkage, resulting in microporosity. The semisolid mushy zone therefore becomes weak and ruptures when stressed in a phenomenon called hot tearing. Although we have not considered fluid flow in the present model, mushy zone solidification behavior is estimated by the solid and residual liquid fractions during coalescence of the $\gamma$ cells.


Diffusion of solute is significantly different in 3D than that in 2D, which results in different velocities, compositions and temperatures of the growing cells/dendrites~\cite{Lee2014,Wang2003}. Lee \emph{et al.}~\cite{Lee2014,Wang2003} have studied the dendrite growth problem in 2D and 3D with and without convection. Their simulations suggest that dendrites are finer in 3D and grow twice as fast in comparison to 2D. Interestingly, the difference in solute mass fraction at the 3D dendrite core with and without convection was less than \SI{0.01}{\%} and the dendrite tip properties such as shape and curvature remained similar in both cases. The topological features of the solid and liquid phase interaction are far more complex in 3D than that in 2D, which makes analysis of the mushy zone difficult. The present work catalogs the above differences that arise from both 2D and 3D simulations in AM solidification conditions.


Solidification in the melt pool begins at the fusion boundary, and $G$ and $V$ are found to vary along this boundary. As in~\cite{Trevor2017}, $G$ and $V$ values are estimated from a 3D heat transfer finite element simulation for use in a phase-field model for microstructure simulation. The phase-field solidification model is described in Sec.~\ref{sec_pf}. Cellular solidification microstructures are presented and analyzed in Sec.~\ref{sec_results}. The general results are discussed in Sec.~\ref{sec_discussion} and conclusions are drawn in Sec.~\ref{sec_conclusions}.

\section{Phase-field solidification model}\label{sec_pf}
The phase-field method is one of the most powerful computational techniques to simulate three-dimensional dendritic growth in binary alloys~\cite{Karma1998,Takaki2014}. We have used a phase-field model that has been detailed in Refs.~\cite{Karma2001,Echebarria2004}. This model quantitatively simulates the time-dependent evolution of a non-conserved phase-field $\phi$ and the conserved composition field $c$ during solidification of a dilute binary alloy. The phase-field $\phi$ is a scalar-valued order parameter field which distinguishes the microstructure phases; $\phi$ = 1 in the solid, $\phi$ = -1 in the liquid and the solid-liquid interface is described by $-1<\phi<1$. This approach avoids explicit tracking of the interface and thus the complex solid-liquid surfaces are extracted in an efficient way \cite{chen2002,boettinger2002,steinbach2009,moelans2008}. An anti-trapping solute flux term was introduced in this model to avoid unphysical solute-trapping effects due to the use of large numerical interface thickness values at low solidification velocities. However, we will show below that the model does not prevent solute-trapping at AM solidification velocities. The effects of melt convection are not included in this model and thus solute is transported in the liquid by diffusion only. The evolution equation for $\phi$ in 3D is expressed as

\begin{eqnarray}\label{eq_phi}
\tau_0 a(\hat{n})^2\frac{\partial \phi}{\partial t} = W_{0}^{2} \nabla \cdot \left[{a(\hat{n})}^2 \nabla\phi\right] + \sum_{i=1}^{3} \partial_i \left[a(\hat{n}) \frac{\partial a(\hat{n})}{\partial(\partial_i \phi)} |\nabla\phi |^2 \right] \nonumber \\
+ \phi -\phi^3 -  \frac{\lambda}{1-k_e} (1-\phi^2)^2 \left[\exp(u) -1 + \frac{G(z-Vt)}{m_l c_0/k_e}\right].
\end{eqnarray}
The dimensionless surface energy function $a(\hat{n}) = 1 - \epsilon \left[3- 4(n_{x}^{4}+n_{y}^{4}+n_{z}^{4})\right]$ represents the three-dimensional fourfold anisotropy at the solid-liquid interface with strength $\epsilon$ and $n_i$ is the interface normal vector pointing into liquid along the Cartesian direction, $i$, in the lab frame of reference. Alloy composition $c_0$, liquidus slope $m_l$, and equilibrium partition coefficient $k_e$ are taken from a Ni-Nb phase diagram~\cite{Nastac1996}. $k_e=c_s/c_l$, where $c_s$ and $c_l$ are the equilibrium compositions on the solid and liquid side of the interface. The dimensionless chemical potential $u$ is given by $\ln \left(\frac{2ck_e/c_0}{1+k_e-(1-k_e)\phi}\right)$. A frozen-temperature approximation is applied in which the temperature gradient $G$ is translated along the $z$ (growth) axis with a velocity $V$.

The evolution equation for $c$ is given by
\begin{equation}\label{eq_c}
\frac{\partial c}{\partial t} = -\nabla \cdot \left[ - \frac{1}{2}(1+\phi)\, D_l \, c \, \exp(u)^{-1}  \, \nabla\exp(u) +\frac{1}{2\sqrt{2}} W_0 (1-k_e) \exp(u) \frac{\partial \phi}{\partial t} \frac{\nabla\phi}{|\nabla \phi|}\right],
\end{equation}
where the first term inside the square bracket represents a standard Fickian diffusion flux and the second term is the anti-trapping solute flux. $D_l$ is the diffusivity of solute in the liquid.

The numerical parameters in this model, $W_0$: the interface thickness, $\tau_0$: the phase-field relaxation time, and $\lambda$: a dimensionless coupling constant, are linked to the material properties via the chemical capillary length $d_0 = a_1 W_{0}/\lambda$ and the time scale for diffusion $\tau_0 = a_2\lambda W_{0}^{2}/D_l$ using a thin-interface analysis which makes the interface kinetics vanish. The numerical constants are given by $a_1$ = 0.8839 and $a_2$ = 0.6267, after Ref.~\cite{Karma2001}. Both $W_0$ and $\tau_0$ values are used to render all the simulation parameters dimensionless.
\subsection{Simulation setup}\label{sec_setup}
In order to study microstructure evolution, $\phi$ (Eq.~(\ref{eq_phi})) and $c$ (Eq.~(\ref{eq_c})) equations of motion are solved on a uniform mesh, using a finite volume method and an explicit time marching scheme. A zero-flux boundary condition is applied on both $\phi$ and $c$ fields in all directions. The size of the simulation box in the growth ($z$) direction is taken as \SI{10}{\micro\metre}, which is at least 50 times the diffusion length $D_l/V$, and varying domain sizes, $L_x \times L_y$, are used ranging from \SI{2}{\micro\metre} $\times$ \SI{2}{\micro\metre} to \SI{4}{\micro\metre} $\times$ \SI{4}{\micro\metre} depending on the fineness of the $\gamma$ cells simulated. For each simulation, a grid spacing $\Delta x/W_0 = \Delta y/W_0 = \Delta z/W_0$ = 0.8 and a time step $\Delta t/\tau_0$ = 0.05 are used. An interface width of $W_0$ = \SI{0.01}{\micro\metre}, corresponding to $\lambda$ = 1.38, is used. 

Each simulation begins from the bottom of the simulation box with a thin solid layer of height \SI{0.05}{\micro\metre} and with an initial Nb composition of $k_ec_0$ in the solid and $c_0$ in the liquid. Small, random amplitude perturbations are applied at the initial solid-liquid interface, from which stable perturbations grow with time and break into steady state $\gamma$ cells. After these cells have grown to a pre-defined length, we translate the box upward and in doing so the solid in the bottom of the box is removed. This approach does not modify the dynamics of the cellular solidification front since there is negligible diffusion in the solid, but saves a considerable amount of computation time and storage. 

\subsection{Parameter details}
It should be noted that we have approximated the Inconel 718 (IN718) multicomponent alloy to be a binary Ni-\SI{5}{\%}~\footnote{Concentration is represented in mass fraction in the present paper.} Nb in this study. The corresponding pseudobinary phase diagram of IN718-\SI{5}{\%} Nb alloy is given in Refs.~\cite{knorovsky1989,Nastac1996}. The solidification sequence of this alloy was found to be completely liquid ($L$) above $T_l = 1637$ K, solid FCC $\gamma$ and $L$ coexistence in $T_s$ $<$ $T$ $<$ $T_l$, and completely solid below $T_s = 1580$ K. $\gamma$-solidification continues until intermetallic eutectic phases, such as Ni$_3$Nb, appear in the microstructure below the eutectic temperature $T_e = 1473$ K. The thermophysical parameters of the dilute IN718-\SI{5}{\%} Nb alloy are taken directly from Refs.~\cite{knorovsky1989,Nastac1996,nie2014} and listed in Tab.~\ref{table_param_pf}. 

\begin{table}[h]
\caption{Material properties used in the simulations, after Refs.~\cite{nie2014,knorovsky1989}.}\label{table_param_pf}
\centering
\begin{tabular}{ll}
\hline
Initial alloy mass fraction, $c_0$ 		&	\SI{5}{\%} \\
Equilibrium partition coefficient, $k_e$	&	0.48	\\
Liquidus slope, $m_l$	&	-10.5 K \%$^{-1}$	\\
Liquid diffusion coefficient, $D_l$	& $3 \times 10^{-9}$ m$^2$ s$^{-1}$ \\
Anisotropy strength, $\epsilon$	& \SI{3}{\%}	\\
Gibbs-Thomson coefficient, $\Gamma$ &  $3.65 \times 10^{-7}$ K m \\
\hline
\end{tabular}
\end{table}

The $\gamma$ cells in a solidifying melt pool grow roughly perpendicular to the solid-liquid interface at a velocity $V$. The solidification parameters ($G$ and $V$ in Eq.~(\ref{eq_phi})) are therefore estimated from the melt pool solid-liquid boundary given by a 3D finite element simulation for a single line laser scan on a single layer IN718 powder with a beam power of 195 W and scan speed of 0.8 m s$^{-1}$, as reported in~\cite{Trevor2017}. $V$ ranges from 0.01 m s$^{-1}$ to 0.3 m s$^{-1}$ and $G$ ranges from $ 2.4 \times 10^7$ K m$^{-1}$ to $0.14 \times 10^7$ K m$^{-1}$ as we move from the bottom to the rear of the solid-liquid boundary in a typical melt pool shape. These $G$ and $V$ values are used to simulate solidification microstructures at different locations within the melt pool.

\section{Results}\label{sec_results}
\subsection{General features of the $\gamma$ phase}
The morphologies in a solidifying melt pool grow in the direction of the temperature gradient $G$ at a velocity $V$. There exist several criteria to determine if the $\gamma$-solidification morphology will be planar, cellular or dendritic. The lower limit for the transition of solidification morphologies can roughly be estimated by satisfying the constitutional supercooling criterion: $V_{cs} = G D_l / \Delta T_0$, where $\Delta T_0 = 57$ K is the freezing range of IN718-\SI{5}{\%} Nb alloy. Whereas the upper limit of the above transition is given by the absolute stability criterion~\cite{Mullins1964}: $V_{ab} = \Delta T_0 D_l / (k_e \Gamma$). The physical meaning of aforementioned limits is that as long as $V$ is below $V_{cs}$, the solidification growth front will be planar, it breaks into cells or dendrites with increasing $V$ above $V_{cs}$, and for $V$ $>$ $V_{ab}$, the morphological instability (Mullins-Sekerka instability~\cite{Mullins1964}) that gives rise to cells/dendrites is suppressed leading back to a planar growth front. Referring to the expressions for $V_{cs}$ and $V_{ab}$, $G$ becomes a less important parameter in the high-velocity limit~\cite{Kurz1994}. We therefore employ a fixed value of $G = 10^7$ K m$^{-1}$ for the phase-field simulations and represent our data as a function of $V$ only. The values of $V$ estimated from FEA simulations are between $V_{cs} < V < V_{ab}$, where $V_{cs}$ = 0.0005 m s$^{-1}$ and $V_{ab}$ = 1 m s$^{-1}$, which suggests that the melt pool will solidify into cellular/dendritic microstructures.

\begin{figure}[h]
\begin{center}
\begin{subfigure}[t]{0.5\textwidth}
\centering
\includegraphics[trim={0cm 0 0 0},clip,height=6cm,width=8cm]{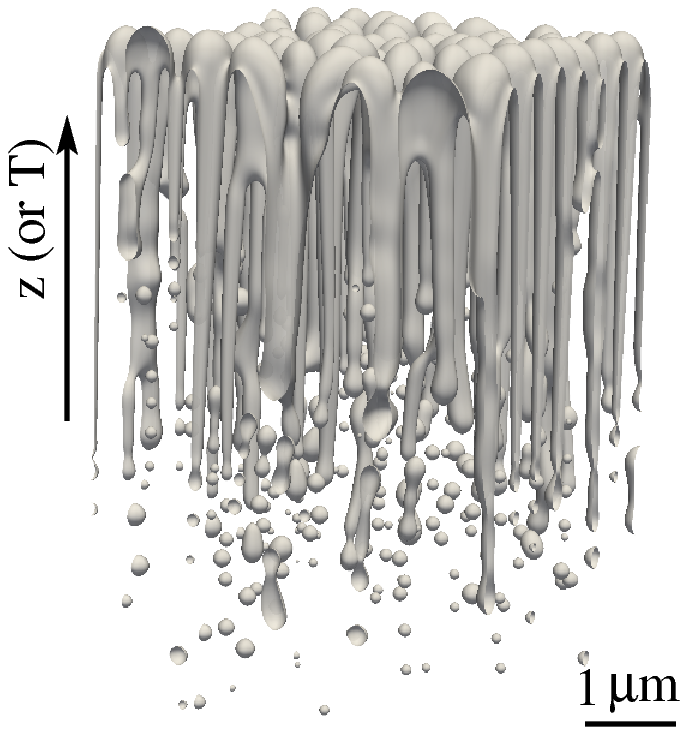}
\caption{}\label{fig_figure1}
\end{subfigure}%
\begin{subfigure}[t]{0.5\textwidth}
\centering
\includegraphics[trim={0cm 4 0 0},clip,height=6cm,width=6cm]{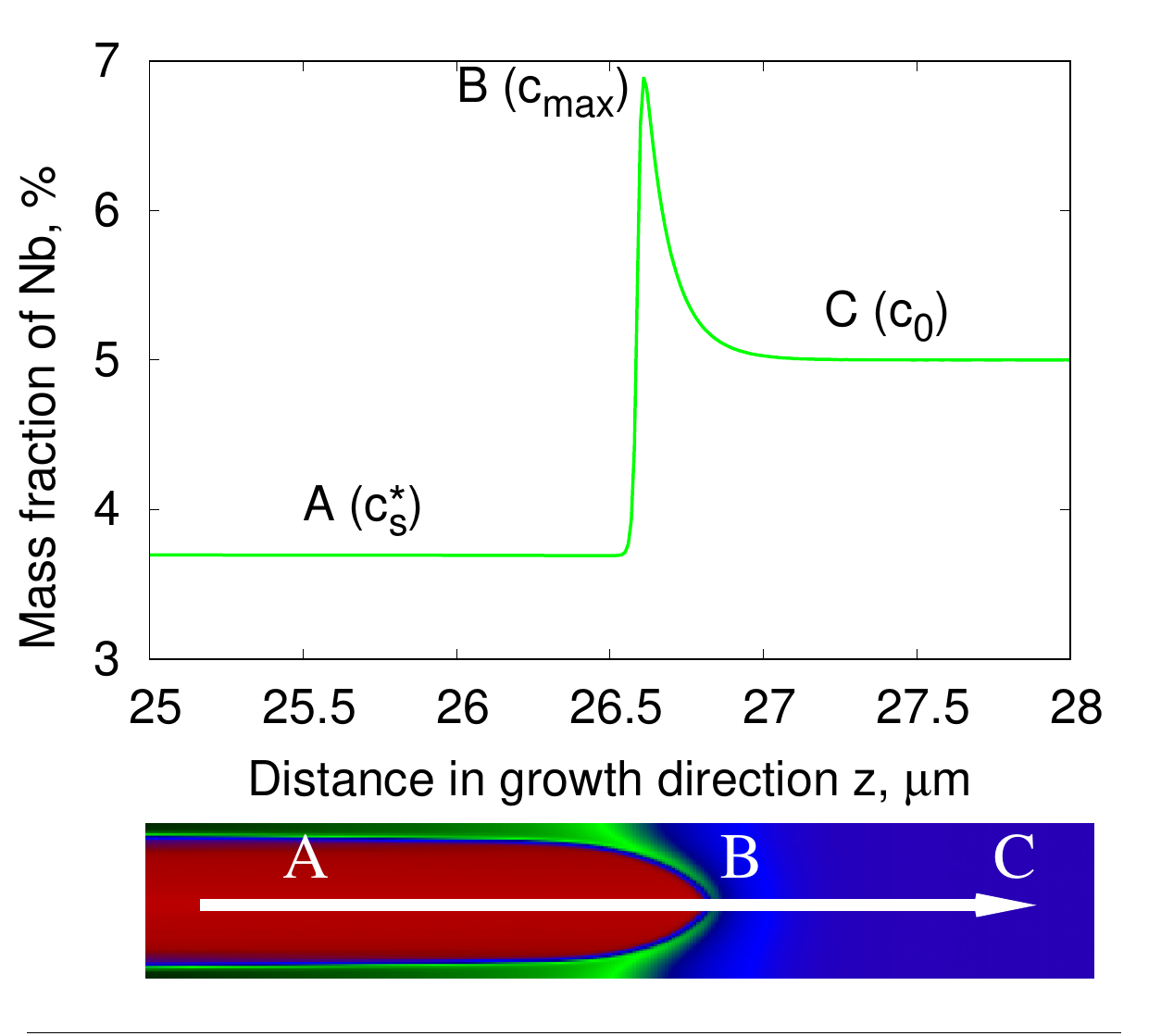}
\caption{}\label{fig_figure2}
\end{subfigure}
\caption{(a) Typical snapshot picture showing a steady state cellular growth front for $V = 0.05$ m s$^{-1}$, which is extracted at the contour $\phi = 0$. Growth direction $z$ is vertical. Average cell spacing $\lambda_c$ measured from this microstructure is \SI{0.22}{\micro\metre}. (b) Variation of Nb along the major axis of a dendrite is shown. Niobium remains constant in the cell core (A) and varies from the cell tip (B) and beyond in the far field liquid (C). The cell core and tip compositions are used to illustrate microsegregation in Sec.~\ref{sec_comp}.}\label{fig_figure}
\end{center}
\end{figure}

The essence of cellular solidification from the phase-field simulations is as follows. Simulations start with the procedure described in Sec.~\ref{sec_setup}.
The initial Mullins-Sekerka instability~\cite{Mullins1964} in the solid-liquid interface is followed by intermediate transient stages of growth by merging or splitting of the neighboring cells, which finally develops in to steady state cellular microstructure. At this stage, the number of cells appearing in the simulation box remains constant and their large-scale geometrical features (growth front, shape and trunk) do not change with time. Cell tips at this stage move with the temperature field at a constant velocity $V$, which equals that estimated from the finite element simulation. We did not notice any quantitative changes in the steady state microstructural features with varying magnitudes of initial interface perturbations in our simulations. The trasient state was extremely short-lived and rapidly reached to steady state. The transient microstructural evolution can further include Mullins-Sekerka instability~\cite{Mullins1964}, which may lead to different transient cellular stages of growth, as discussed in \cite{Warren1990,Warren1993}. The present work focuses on the analysis of well developed $\gamma$-cells when transient effects are absent. A typical example of the simulated 3D cells is presented in Fig.~\ref{fig_figure}.

Three essential features can be seen in the cellular microstructure of Fig.~\ref{fig_figure1}. First, the average distance between the tips of neighboring $\gamma$-cells remains constant at steady state, which we refer to henceforth as the primary cell spacing $\lambda_c$. In our simulations, $\lambda_c$ varies with $V$, which is described in Sec.~\ref{sec_spacing}. Second, Nb composition varies between the liquid ahead of the cell tips and the intercellular liquid, and along the cells (Fig.~\ref{fig_figure2}). An analysis of the Nb concentration field is given in Sec.~\ref{sec_comp}.

Third, spherical droplets feature in the cell roots. In order to maintain the steady state average distance between the cell tips and the cell roots, Nb-rich droplets periodically pinch off from the bottom of intercellular liquid, resulting in a discontinuous array of spherical pockets in the $\gamma$ solid. These Nb-rich droplets could transform to secondary eutectic phases below the eutectic temperature. Similar microstructural features have also been reported in experiments as well as in simulations~\cite{ungar1985,Boettinger1988,Boettinger1999}. 

\subsection{Characteristic $\gamma$ cell spacings}\label{sec_spacing}
\begin{figure}[h]
\begin{center}
\includegraphics[scale=0.75]{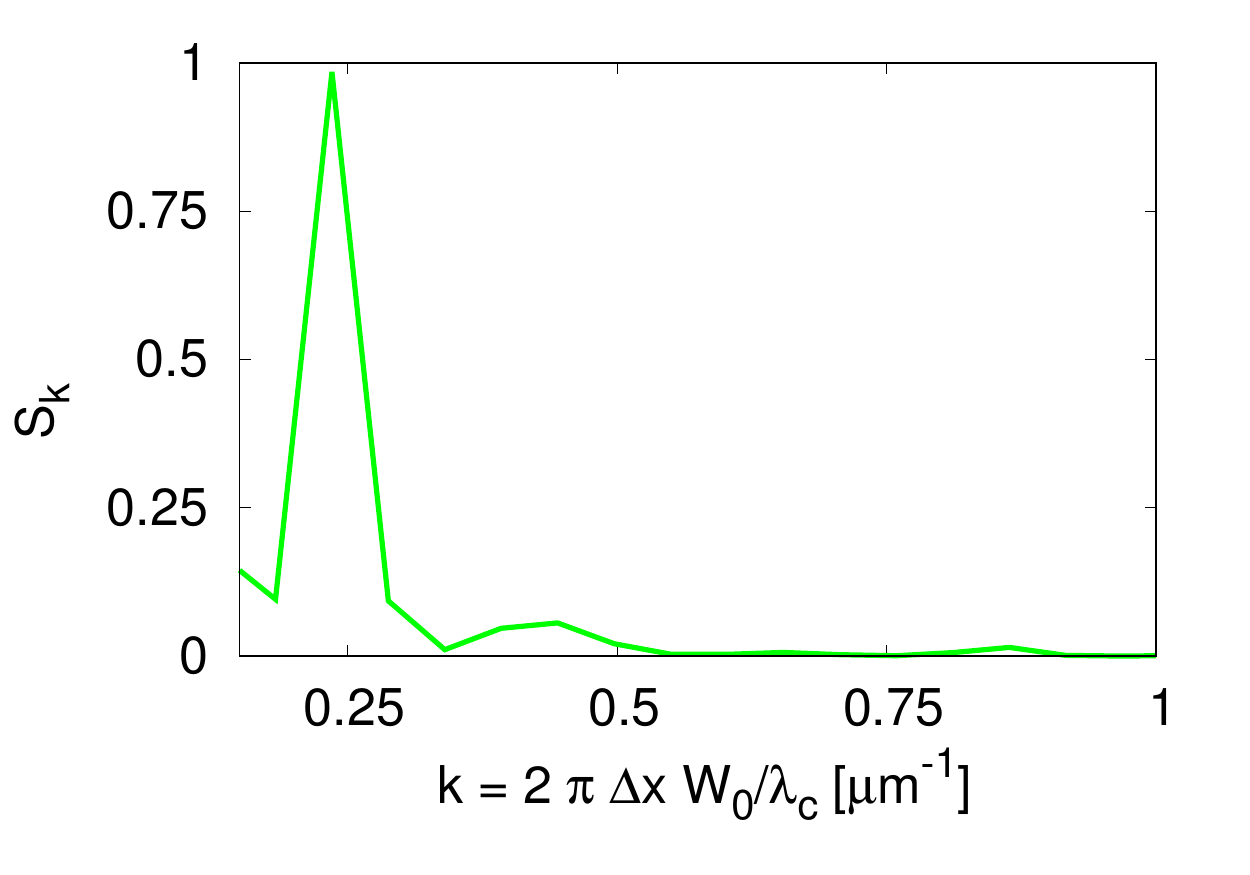}
\caption{Power spectrum calculated from a simulated 3D cellular microstructure for $V$ = 0.05 m s$^{-1}$. The main peak corresponds to the primary cell spacing $\lambda_c$ in the microstructure. Note that we consider the microstructure only above the solidus temperature $T_s$ above which no droplets are seen.}\label{fig_sf}
\end{center}
\end{figure}

The average cell spacing $\lambda_c$ developed in the simulated 3D cellular microstructure is estimated by calculating the mean power spectrum $S_k = |h(k)|^2$, where $h(k)$ is the Fourier transform of the solid-liquid surface profile $h(z)$ and $k$ is the wave number, in the following manner 

\begin{equation}
\frac{2\pi}{\lambda_c} = k_{mean} = \frac{\sum_{k>0} kS_k}{\sum _{k>0} S_k}.
\end{equation}
We show one instance of such a spectrum in Fig.~\ref{fig_sf}. The main peak in this spectrum corresponds to the dominant wavelength in the microstructure, i.e. $\lambda_c$. For the melt pool solidification conditions considered, $\lambda_c$ obtained in 3D simulations ranges from \SI{0.1}{\micro\metre} to \SI{0.5}{\micro\metre}, while in 2D simulations $\lambda_c$ ranges from \SI{0.2}{\micro\metre} to \SI{1.1}{\micro\metre}. On average $\lambda_c$ is smaller in 3D than that in 2D, since solute diffusion at the cell tips is more efficient in 3D than that in 2D simulations. Similar cell spacing data have also been reported in recent L-PBF experiments~\cite{Amato2012,Murr2012} conducted on an Inconel alloy with the same laser processing parameters used in~\cite{Trevor2017}.

\begin{figure}[h]
\begin{center}
\includegraphics[scale=0.7]{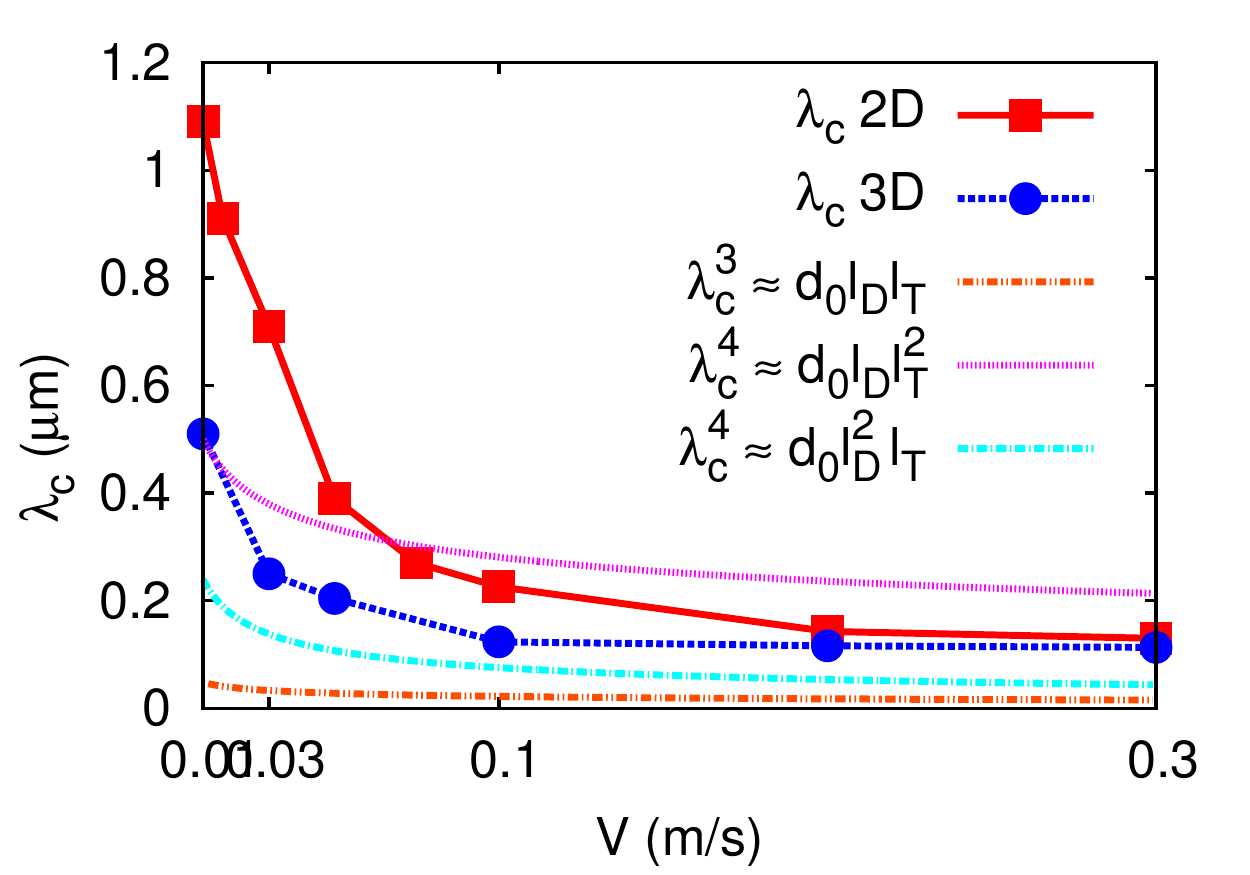}
\caption{Phase-field simulated cell spacing data are compared with different scaling predictions. 3D data are best approximated between $(d_0l_D^2l_T)^{1/4}$ and $(d_0l_Dl_T^2)^{1/4}$.}\label{fig_spacing}
\end{center}
\end{figure}

Characteristic steady state spacings that develop in the solidification microstructures are generally a geometric mean of three different length scales associated during the growth of interfacial morphological instabilities into cellular structures~\cite{Rappazbook,Mullerbook,Trivedi1994}. These lengths are the capillary length $d_0$, thermal length $l_T = \Delta T/G$, and diffusion length $l_D = D/V$. The average cell spacing thus can be represented by a product mixture of these length scales resulting in power laws of the form $(d_0l_Dl_T)^{1/3}$, $(d_0l_D^2l_T)^{1/4}$ or $(d_0l_Dl_T^2)^{1/4}$, as reviewed in \cite{Rappazbook}. A comparison of phase-field spacing data with these predictions is shown in Fig.~\ref{fig_spacing}. It is evident that the data matches none of the predictions, however, on average they are between $(d_0l_D^2l_T)^{1/4}$ and $(d_0l_Dl_T^2)^{1/4}$, or $\propto G^{-m}V^{-n}$, with $m$ and $n$ between 0.25 to 0.5, when expressed in terms of the solidification conditions $G$ and $V$.

Different analytical models are available to estimate $\lambda_c$ from a cellular/dendritic microstructure. To mention a few, the model of Hunt and Lu \cite{LuandHunt1992,HuntandLu1993} predicts $\lambda_c \propto V^{-0.6}$ with no significant dependence on $G$. Ma and Sham \cite{Ma1998} estimated $\lambda_c \propto V^{-0.5}$ also with no dependence on $G$. While independent works of Hunt~\cite{hunt1979}, Trivedi~\cite{Trivedi1984} and Kurz and Fisher~\cite{kurz1986} estimated $\lambda_c \propto G^{-0.5}V^{-0.25}$, although from different approaches. However, none of these correlations predict the spacing selection in arbitrary solidification conditions~\cite{Whitesell2000}. We will show below that cell growth variables such as dendrite tip radius, tip temperature and tip concentration vary with $V$ (also reviewed in Ref.~\cite{Wang2011}) and therefore the above theoretical estimations for $\lambda_c$ based on any tip operating conditions and geometrical approximations (ellipsoid, hemispherical) for the cell shape do not strictly apply~\cite{kurzbook,Rappazbook}.
\subsection{Characteristic $\gamma$ solid fractions}
\begin{figure}[h]
\begin{center}
\includegraphics[scale=0.8]{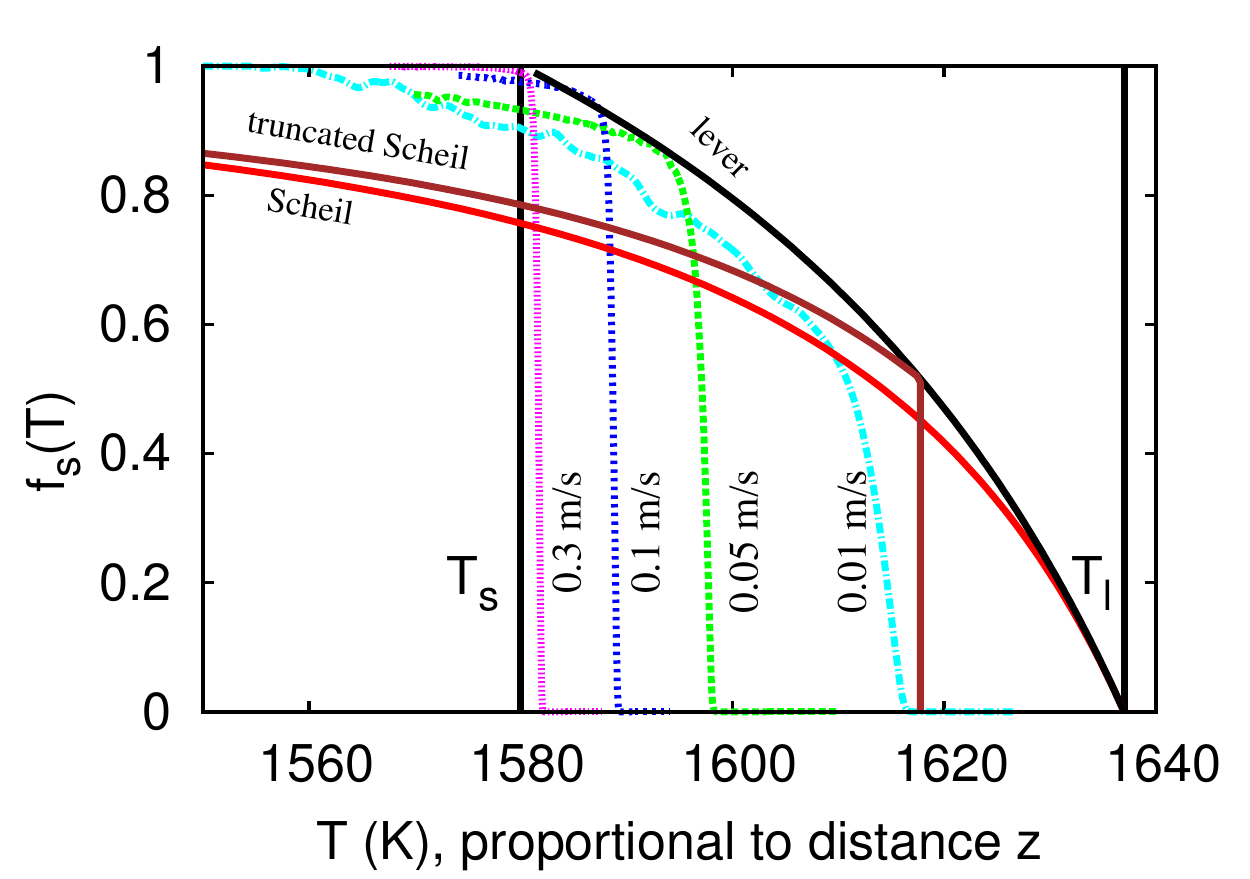}
\caption{Solid fraction $f_s$ vs. $T$ is presented for various $V$. $f_s$ (or $f_l$ ) decreases (increases) with increasing $T$. $f_s$ is 0 in the liquid and approaches 1 during terminal solidification. $T_s$ and $T_l$ are shown for illustration. The simulation data are compared with the Scheil, lever and truncated Scheil approximations for reference.}\label{fig_solidfraction}
\end{center}
\end{figure}

The solidification paths of the simulated cellular microstructures are approximated by extracting the data for volume fraction of solid $f_s$ as a function of temperature $T$, i.e. $f_s(T)$. Note that $T$ is proportional to the distance in growth direction $z$. $f_s(T)$ is estimated by the area fractions of the solid in the $xy$-planes perpendicular to the growth axis $z$ as
\begin{equation}\label{eq_solidfraction}
f_s(T(z)) = \frac{1}{L_x L_y} \int \frac{1}{2} \left[ 1 + \phi(x,y,z) \right] dx\, dy,
\end{equation}
where $\phi$ = 1 in the solid and -1 in the liquid. The calculation of $f_s(T)$ is illustrated in Fig.~\ref{fig_solidfraction}. $f_s(T)$ is equal to zero in the liquid ahead of the cell front and increases behind the cell front as a function of depth in the mushy region until $f_s(T) \rightarrow 1$ during late stages of solidification close to bottom of the simulation box. The solid-liquid interface close to the cellular growth front can roughly be approximated by the $f_s(T)$ curves near $f_s$ = 0. The steepness of the $f_s(T)$ curves, i.e. $|df_s/dT|$, decreases with decreasing $V$. This indicates that the intercellular liquid regions become deeper with decreasing $V$ (from $V_{ab}$) and thus liquid can exist up to greater depths in between the cells. A measurement of $|df_s/dT|$ near $f_s = 0$ is therefore used to describe the distribution of liquid in the intercellular channels that influence other processes, such as the solidification shrinkage and cracking behavior during late stages of solidification~\cite{Kou2015,Provatas2017,Rappaz1999}. A longer liquid channel hinders feeding of the liquid to the bottom of intercellular regions and thus affects solidification shrinkage and promotes solidification crack formation in the mushy zone. In our simulations, $|df_s/dT|$ ranges from 0.1 K$^{-1}$ to 1.4 K$^{-1}$ as $V$ ranges from 0.01 m s$^{-1}$ to 0.3 m s$^{-1}$. Although the present simulations are conducted for a constant $G$ and $c_0$, $f_s(T)$ is a function of $G$ and $c_0$, since the length ($\approx$ $\Delta T_0/G$) of intercellular liquid channels varies with these parameters~\cite{kurzbook,Rappazbook,Provatas2017}.

Referring to Fig.~\ref{fig_solidfraction}, the length of the mushy zone is largest for low $V$ and decreases with increasing $V$, which suggests that the probability of defect formation is also higher at low $V$. The statistics of the mushy zone in $f_s(T)$ curves are noisy which is due to rapid connection and disconnection events between the cell roots at different depths within the mushy zone leading to non-uniform distribution of solid and liquid phases. Recent studies in Refs.~\cite{Ma2014,Space2001} have also found such non-steady behavior in the mushy zone, resulting in similar $f_s(T)$ curves as ours. We calculate $f_s(T)$ using the Scheil equation~\cite{kurzbook,Rappazbook}, the lever rule~\cite{kurzbook,Rappazbook} and the truncated Scheil approximation~\cite{Flood1987} for reference. The discrepancies between the microsegregation models and our phase-field simulations are evident in Fig.~\ref{fig_solidfraction}. The Scheil equation assumes no diffusion in the solid while the lever rule assumes infinite diffusion in the solid for a planar solid-liquid interface. The truncated Scheil solid fraction is shown for $V$ = 0.01 m s$^{-1}$, which takes into account the cell tip composition (from Eq.~(\ref{eq_iv})) to calculate $f_s$ and is close to the simulation data for 0.01 m s$^{-1}$. The $f_s(T)$ curves are used in the following sections to extract the cell tip temperatures as well as the bridging temperatures when the cell roots grow toward each other and coalesce in the mushy zone. The mushy zone coalescence behavior is further elaborated in Sec.~\ref{sec_temp}.

\subsection{Characteristic compositions: microsegregation, solute-trapping}\label{sec_comp}
\begin{figure}[h]
\begin{center}
\includegraphics[scale=0.8]{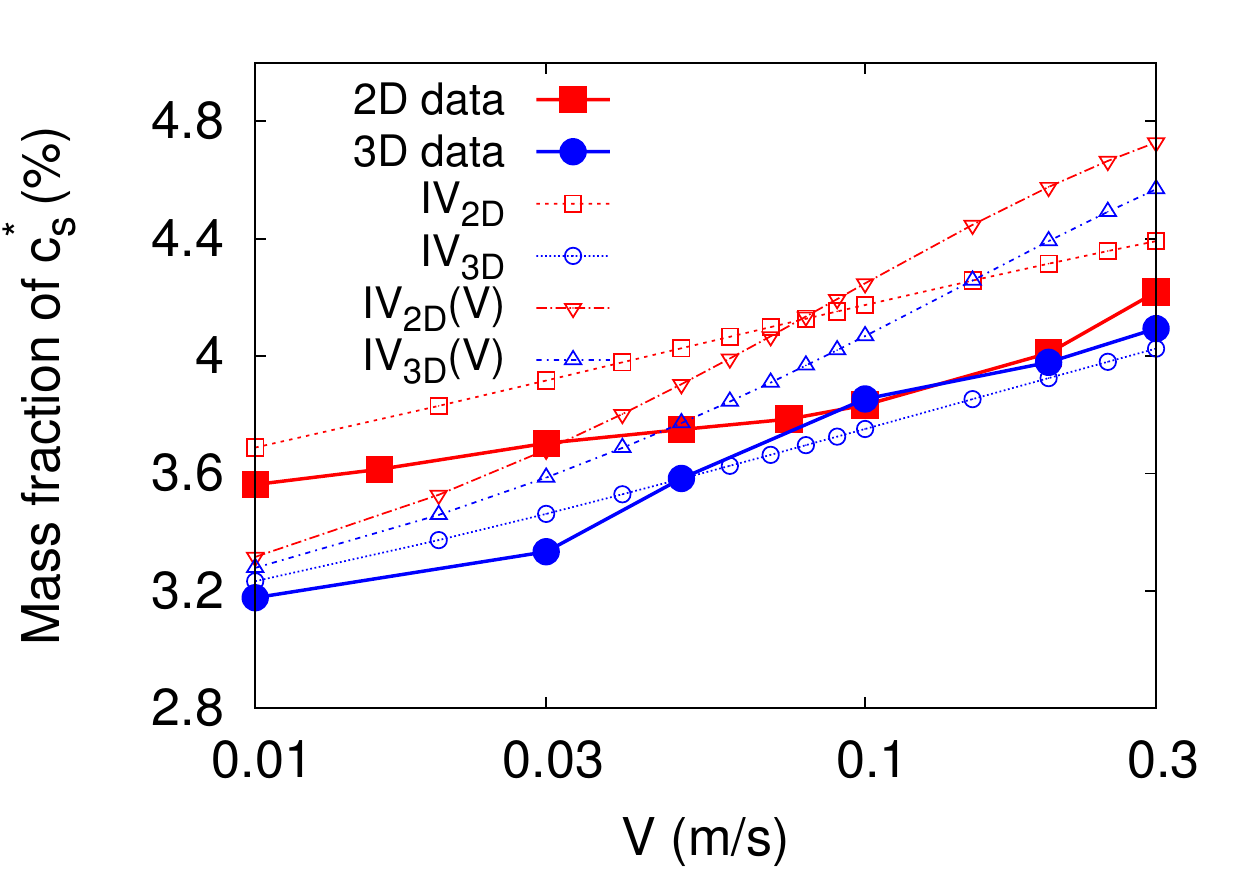}
\caption{3D data and predictions are represented by blue and 2D data and predictions are represented by red. Simulated cell core composition $c_{s}^{*}$ values are compared with the theory considering equilibrium partitioning at the solid-liquid interface given by $k_e$ (Eq.~(\ref{eq_iv})). Equation~(\ref{eq_iv3d}) is substituted in Eq.~(\ref{eq_iv}) to obtain 3D prediction for $c_{s}^{*}$. Equation \ref{eq_iv2d} is substituted in Eq.~(\ref{eq_iv}) to obtain 2D prediction for $c_{s}^{*}$. While 3D data are close to the prediction, 2D data are far from this. The comparison with $V$-dependent solute partitioning coefficient, given by $k_v$, is also presented in order to illustrate the degree of non-equilibrium solute-trapping behavior. For details, please refer to the text.}\label{fig_iv}
\end{center}
\end{figure}

Microsegregation (i.e. spatial distributions of Nb in the solidified $\gamma$ phase) is inevitable during AM solidification processing. Niobium composition varies from the cell core ($c_s^{*}$) to the cell tip ($c_{\mbox{\scriptsize max}}$) and beyond in the far field liquid ($c_0$) (Fig.~\ref{fig_figure2}). $c_s^{*}$ remains nearly constant inside the cells as the diffusivity of solid is negligible. $c_s^{*}$ values are extracted from the simulations for each $V$, using a procedure given in Ref.~\cite{amberg2008}, and plotted in Fig.~\ref{fig_iv}. $c_{s}^{*}$ increases with increasing solid-liquid interface velocity $V$ due to insufficient time for Nb to diffuse from the solid to the liquid. $c_{s}^{*}$ can be estimated following Kurz and Fisher~\cite{kurzbook} who analyzed the diffusion fields around an isolated cell/dendrite tip of paraboloid (3D) or parabolic (2D) revolution with zero capillarity which led to the following analytical equations
\begin{eqnarray}
G_c &=& -\frac{V}{D_l}\frac{c_{s}^{*}}{k_e}(1-k_e), \label{eq_iv_first} \\
R &=& 2\pi\left[ \frac{\Gamma}{m_l G_c - G} \right]^{\frac{1}{2}}, \\
c_{s}^{*} &=& \frac{k_e c_0}{1-(1-k_e)Iv(P)}, \label{eq_iv} \\
Iv(P) &\equiv& P \exp(P) E_1(P) \;\;\mbox{(in 3D)}, \label{eq_iv3d} \\
Iv(P) &\equiv& \sqrt{\pi P} \exp(P) \, \text{erfc}(\sqrt{P}) \;\;\mbox{(in 2D)}, \label{eq_iv2d} \\
P &=& \frac{RV}{2D_l},  \label{eq_iv_last} \\
\Delta T &=& -\frac{m_l c_0 Iv(P) (1-k_e)}{1-(1-k_e)Iv(P)} + \frac{2\Gamma}{R}.  \label{eq_undercooling}
\end{eqnarray}
In the above equations, $G_c$ is the composition gradient in the liquid, $R$ the cell/dendrite tip radius, $c_{s}^{*}$ the cell core composition, $E_1(P)$ the first exponential integral of the cell/dendrite P\'{e}clet number $P$ and $\Delta T$ the cell tip undercooling. The expressions for the Ivantsov function $Iv(P)$ differ in 2D and 3D to represent different cell tip geometries. We compare our simulation results with the above theory in Fig.~\ref{fig_iv}. The comparison between the 3D data and the theory is reasonably good with an error on the order of $\pm$ \SI{5}{\%}, while the 2D data are far from the prediction. Note that 3D simulations predict a lower Nb concentration in the cell core since more Nb can diffuse in/out from a 3D cell compared to a 2D cell.

\begin{figure}[h]
\begin{center}
\includegraphics[scale=0.75]{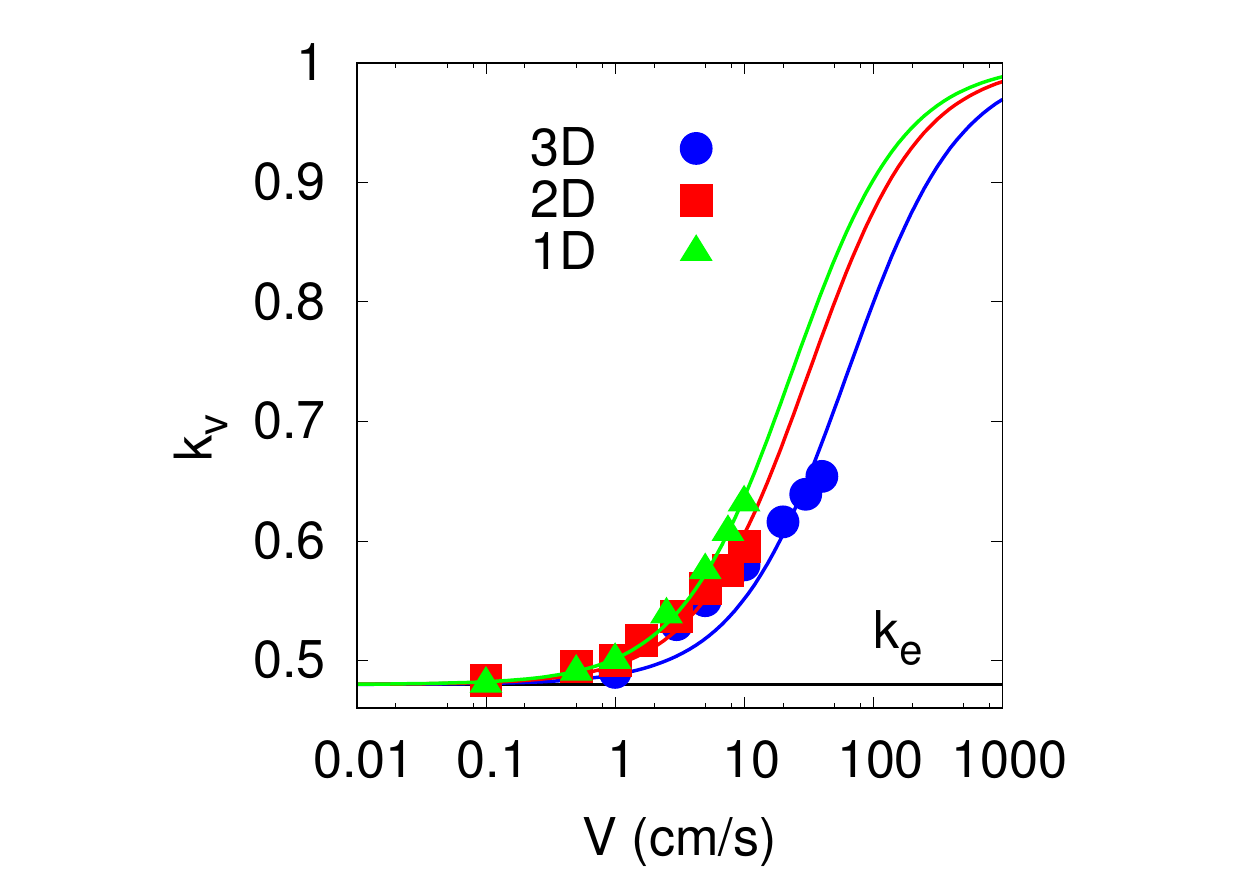}
\caption{The extracted $k_v$ values from 3D simulations are plotted for various $V$ along with a fit to the Aziz function \cite{Aziz1982}. 2D and 1D simulation data are also presented using the same approach to illustrate the effect of dimensionality on the partitioning of solute at the solid-liquid interface. Note that with increasing $V$, $k_v$ increasingly deviates from equilibrium $k_e$.}\label{fig_kv}
\end{center}
\end{figure}

Close to the cell tips, there is a spike in Nb composition $c_{\mbox{\scriptsize max}}$ due to rejection of Nb by the growing cells (Fig.~\ref{fig_figure2}). Although not shown here, $c_{\mbox{\scriptsize max}}$ varies with $V$ in a manner similar to $c_s^{*}$. Such $V$-dependent solute partitioning indicates that the equilibrium partition coefficient, $k_e$, is not recovered at the solid-liquid interface and thus the local interface equilibrium fails. The partitioning of Nb in the solid and liquid is thus described by a $V$-dependent microsegregation coefficient $k_v$~\cite{Aziz1982}, which reads as $k_v (V) = \frac{c_{s}^{*}}{c_{\mbox{\scriptsize max}}}$. The $k_v$ data are plotted against $V$ and fitted with the Aziz solute-trapping function~\cite{Aziz1982}, given by $k_v (V) = \frac{k_e + V/V_D}{1 + V/V_D}$, to obtain a characteristic solute-trapping velocity $V_D$ = 62 cm s$^{-1}$ (Fig.~\ref{fig_kv}). Note that solute partitioning across a 2D cellular growth front resulted in $V_D$ = 32 cm s$^{-1}$, whereas across a planar growth front $V_D$ = 23 cm s$^{-1}$. The extent of solute partitioning is different in different dimensions, primarily due to the average curvature at the cell tips, as illustrated in the literature~\cite{Rappazbook}.

In order to predict the dependence of the extent of cell core microsegregation on $V$, the theoretical predictions are made $V$-dependent by replacing $k_e$ with $k_v$ in Eqs.~(\ref{eq_iv_first})--(\ref{eq_undercooling}) in a heuristic approach. The $V$-dependent prediction for  $c_s^{*}$ following $k_v (V)$ values is given in Fig.~\ref{fig_iv}. The differences in $c_s^{*}$ between the theory (using $k_v$) and simulated values (using $k_e$) at each $V$ predict the range of solute-trapping, the magnitude of which increases with increasing $V$. This behavior is expected when the AM velocity approaches $V_{ab}$. Similar solute trapping behavior has been reported in recent numerical studies~\cite{Boettinger1999,amberg2008,Lee2010,Kundin2015}. Note that the effect of $k_v$ is included in Eqs.~(\ref{eq_iv_first})--(\ref{eq_undercooling}) in an ad hoc manner. Therefore, we may be neglecting important terms or features that might otherwise be present in a formal, more self-consistent derivation of the velocity dependence. If the curvature correction is included in $k_e$ following $k_e(1-(1-k_e)d_0/\rho)$, where $\rho$ is the cell tip curvature, the predictions for $c_s^{*}$ become closer to the simulation data in Ref.~\cite{Mullis2010}. Moreover, the dendrite growth theory does not include the anisotropy effects associated with the surface energy. The anti-trapping solute flux in Eq.~(\ref{eq_c}) needs to be reformulated in order to match solute trapping behavior in real alloys during AM processes.

\subsection{Characteristic temperatures: tip and coalescence}\label{sec_temp}
\begin{figure}[h]
\begin{center}
\includegraphics[scale=0.75]{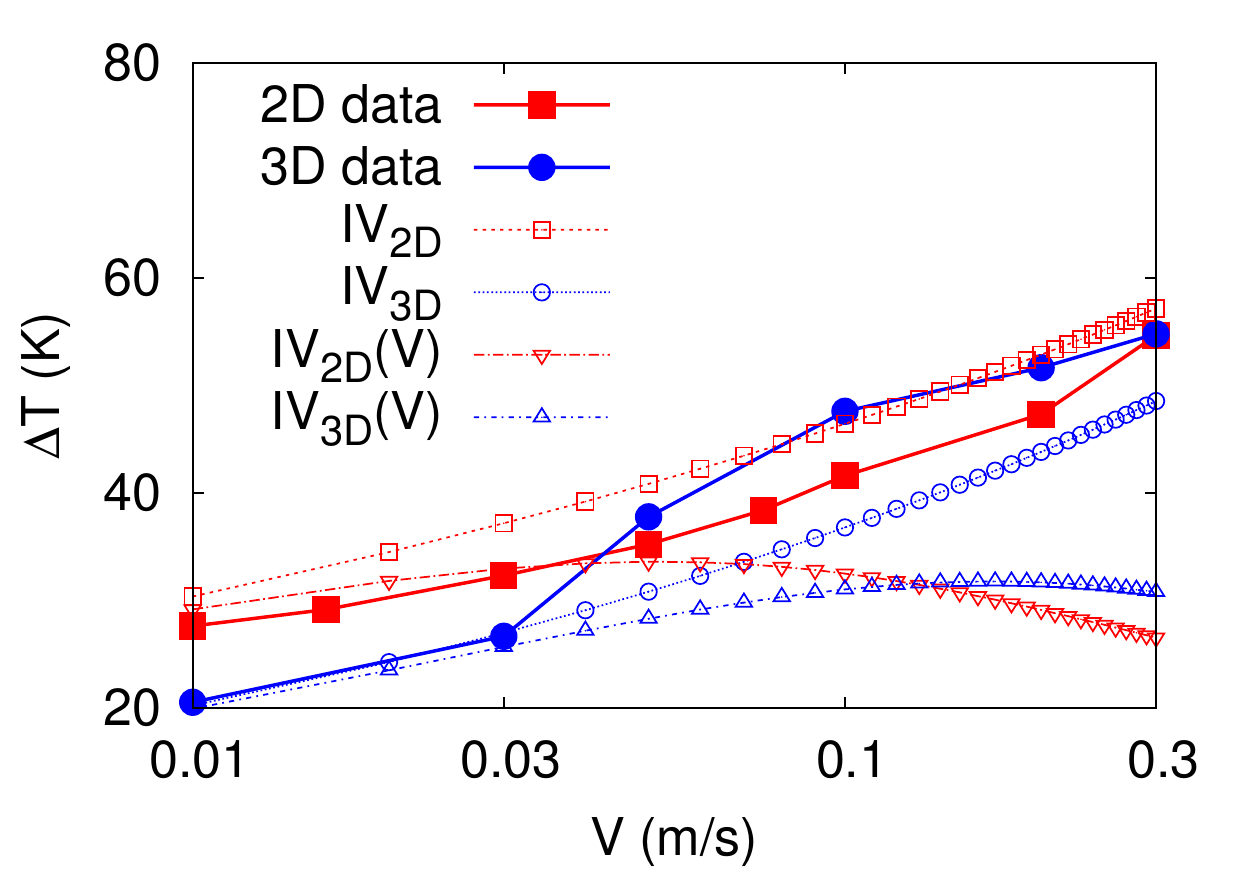}
\caption{3D data and predictions are represented by blue and 2D data and predictions are represented by red. Cell tip undercooling $\Delta T$ increases with $V$. 3D and 2D data are compared with theoretical estimates given by Eq.~(\ref{eq_undercooling}) for constant $k_e$. The comparison with V-dependent $k_v$ is also presented to illustrate deviation from the interface equilibrium. For details, please refer to the text.}\label{fig_undercooling}
\end{center}
\end{figure}

The determination of the cell tip temperature is important as it controls the Nb partitioning in the tip region of the growing cell and hence the final microsegregation pattern in the $\gamma$ solid. The cell tip temperature $T_{\mbox{\scriptsize tip}}$ or the tip undercooling $\Delta T$ below $T_l$ is predicted by Eq.~(\ref{eq_undercooling}). The prediction is then compared with the phase-field data in Fig.~\ref{fig_undercooling}. The $\Delta T$ curves suggest that $\Delta T$ increases (or $T_{\mbox{\scriptsize tip}}$ decreases) with increasing $V$. These results are consistent with Fig.~\ref{fig_solidfraction}, where cell tips (approximated by $f_s \rightarrow 0$) are at lower temperatures with increasing $V$. The $\Delta T$ data and theory (using $k_e$) are comparable in Fig.~\ref{fig_undercooling}. When the prediction for $\Delta T$ is made using the $V$-dependent parameters, the comparison deviates at high $V$ approaching $V_{ab}$. Note that the diffusion of heat is ignored in the present scenario, and a linear frozen description is used to represent temperature. As our data suggest, these approximations may not accurately reflect the AM solidification regime.

\begin{figure}[h]
\begin{center}
\begin{subfigure}[t]{0.3\textwidth}
\begin{center}
\includegraphics[height=3cm,width=3cm]{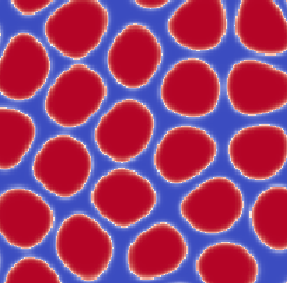}
\caption{}\label{fig_plane3}
\end{center}
\end{subfigure}%
\begin{subfigure}[t]{0.3\textwidth}
\begin{center}
\includegraphics[height=3cm,width=3cm]{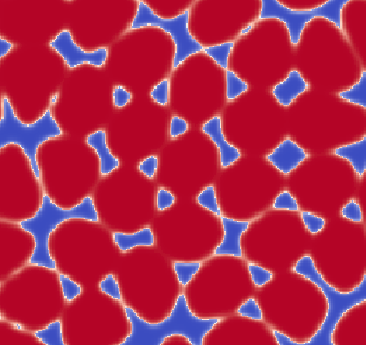}
\caption{}\label{fig_plane2}
\end{center}
\end{subfigure}%
\begin{subfigure}[t]{0.3\textwidth}
\begin{center}
\includegraphics[height=3cm,width=3cm]{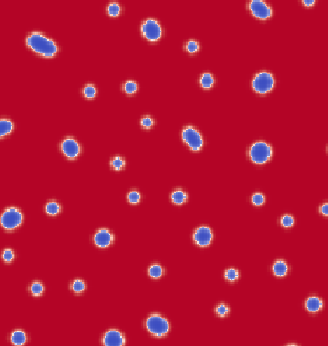}
\caption{}\label{fig_plane1}
\end{center}
\end{subfigure}
\caption{Typical snapshot pictures showing connections between the cellular microstructural features varies with depth $z$ (or $T$). Blue represents liquid and red represents solid. (a) Solid is disconncted and liquid is connected in a plane in the liquid. (b) Solid cells and liquid channels coalesce in the mushy zone. (c) Solid is connected and liquid is disconnected in a plane deep in the solid.}\label{fig_plane}
\end{center}
\end{figure}

\begin{figure}[h]
\begin{center}
\includegraphics[scale=0.75]{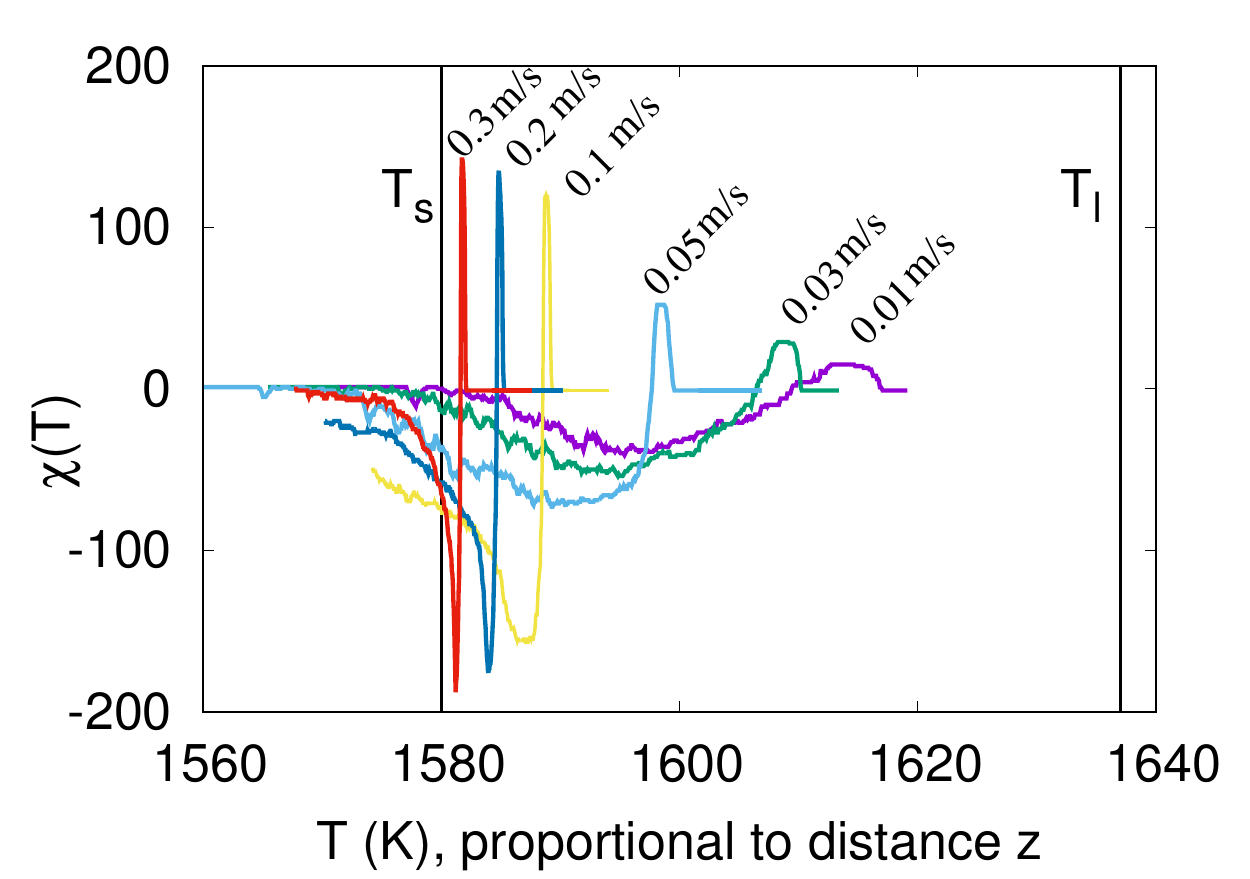}
\caption{Euler characteristic $\chi(T)$ is presented for various $V$. Solid is disconnected and liquid is connected close to the cell tips when $\chi$ is positive. In contrast, solid is connected and liquid is disconnected when $\chi$ is negative. The maximum of $\chi$ roughly represents the number of cells in the simulation box. In this approach, a plane in the mushy zone represented by $\chi$ = 0 is the bridging plane.}\label{fig_connection}
\end{center}
\end{figure}

Behind the cell tips in the mushy zone, there is a characteristic plane where the neighboring cell roots coalesce with each other, separating the liquid into isolated droplets. An analysis of this plane in the AM solidification regime is essential as the temperature and liquid fraction of this plane influence the tensile strength and resistance to rupture of the semisolid mushy zone. We represent the characteristic bridging plane of the mushy zone by an isotherm $T_{\mbox{\scriptsize bridge}}$. It is difficult to extract $T_{\mbox{\scriptsize bridge}}$ from the root of the cells due to apparent randomness in forming the connections and disconnections between the cell roots at regular time intervals leading to bridge formations and droplet pinch offs. A measure of the Euler characteristic is therefore used to represent the coalescence plane. In order to extract $T_{\mbox{\scriptsize bridge}}$ from the 3D mushy zone, we use the Hoshen-Kopelman algorithm, as given in Ref.~\cite{Hoshenkopelman}. The algorithm measures connectivity between the mathematical sets (or clusters) of a particular field, such as the phase-field $\phi$. The algorithm returns the Euler characteristic $\chi$ in terms of the difference between the number of solid and liquid clusters in a $xy$-plane. A brief procedure for the implementation of the algorithm is given here. First we convert the $\phi$-field of a $xy$-plane to either 1 or -1 depending on the sign of $\phi$ at each grid point and then count the number $N$ of connected and/or disconnected sets of 1 or -1. The Euler characteristic of the $xy$-plane is then defined by $\chi$ = N$\phi_{+1}$ - N$\phi_{-1}$. Close to top of the simulation box, where no cells and only liquid is present, the number of liquid clusters equals to 1 and the number of solid clusters equals to zero, making $\chi = -1$. With increasing distance from the top, solid clusters appear disconnected in a $xy$-plane close to the tip region (Fig.~\ref{fig_plane3}). With further increasing distance from the top, solid clusters begin to connect with each other and thus liquid becomes isolated in between the solids (Fig.~\ref{fig_plane2}). Whereas in a plane deep in the solid, solid is continuous and liquid is disconnected (Fig.~\ref{fig_plane1}). Referring to the above planes, the variation of $\chi$ is shown in Fig.~\ref{fig_connection}. The value of $\chi$ changes from positive to negative as the number of isolated liquid clusters becomes dominant deep in the coalesced solid. A value of $\chi$ = 0 in the mushy zone represents a plane where solid cells and liquid channels are very connected, that is, the bridging plane (Fig.~\ref{fig_plane2}).

In Fig.~\ref{fig_connection}, the value of $\chi$ = -1 in the far field liquid suggests that there are no cells and only liquid is present. With increasing distance from the liquid, there is a spike in $\chi$, which roughly scales with the number of cells in the simulation box. As we further move towards the solid, the liquid channels become highly disconnected and $\chi$ changes sign from positive to negative. The $\chi(T)$ data are noisy in the mushy zone which is not surprising due to the apparent randomness during coalescence and pinch off events close to the cell roots. We do not smooth these curves, as our objective is to extract the value of $T$ when $\chi$ vanishes only.

\begin{figure}[h]
\begin{center}
\includegraphics[scale=0.75]{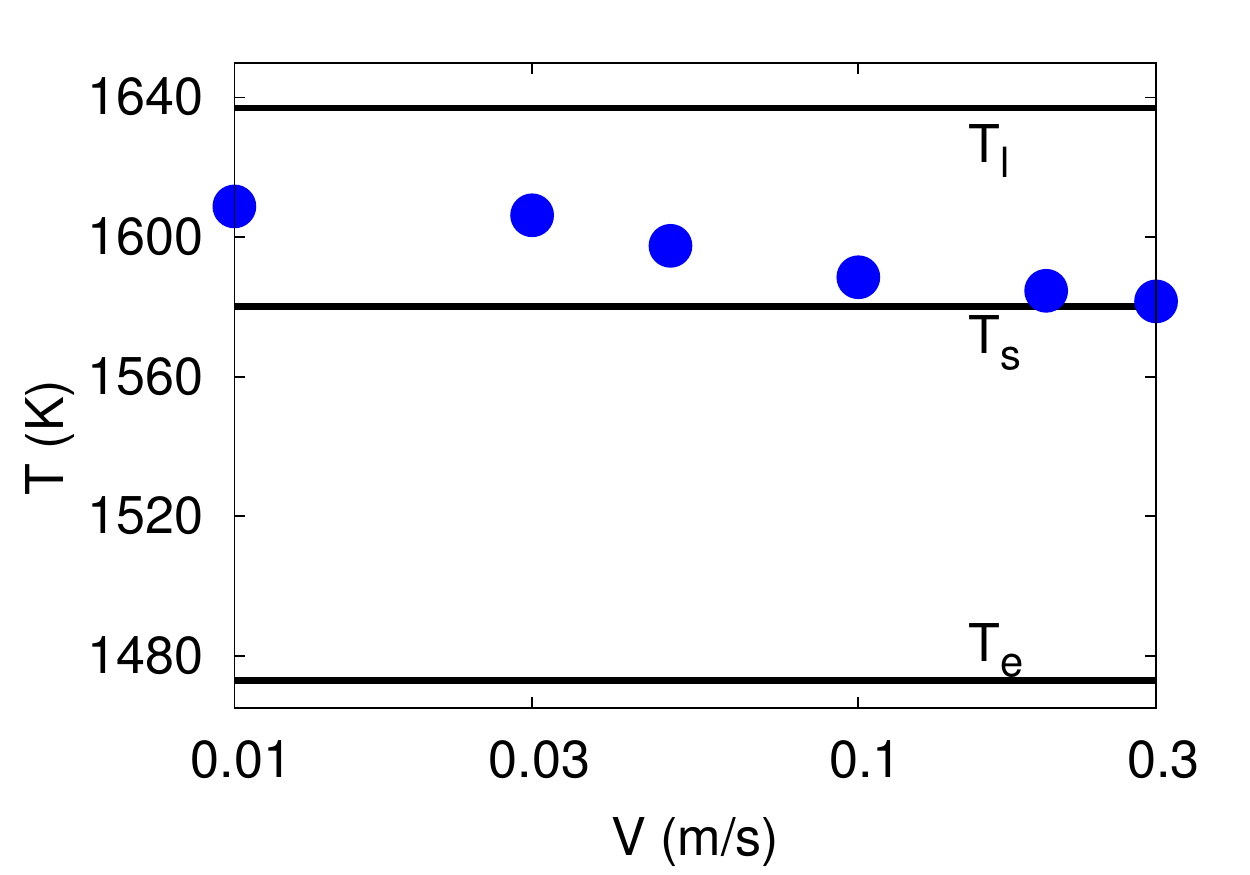}
\caption{The circular data points represent $T_{\mbox{\scriptsize bridge}}$, which decreases with increasing $V$. This signifies that the last remaining liquid in the mushy zone exist at lower temperatures when $V$ is lowered. $T_s$, $T_l$ and $T_e$ are used to illustrate $T_{\mbox{\scriptsize bridge}}$ relative to different invariant temperatures in the phase diagram \cite{knorovsky1989,Nastac1996}.}\label{fig_tempb}
\end{center}
\end{figure}

\begin{figure}[h]
\begin{center}
\includegraphics[scale=0.75]{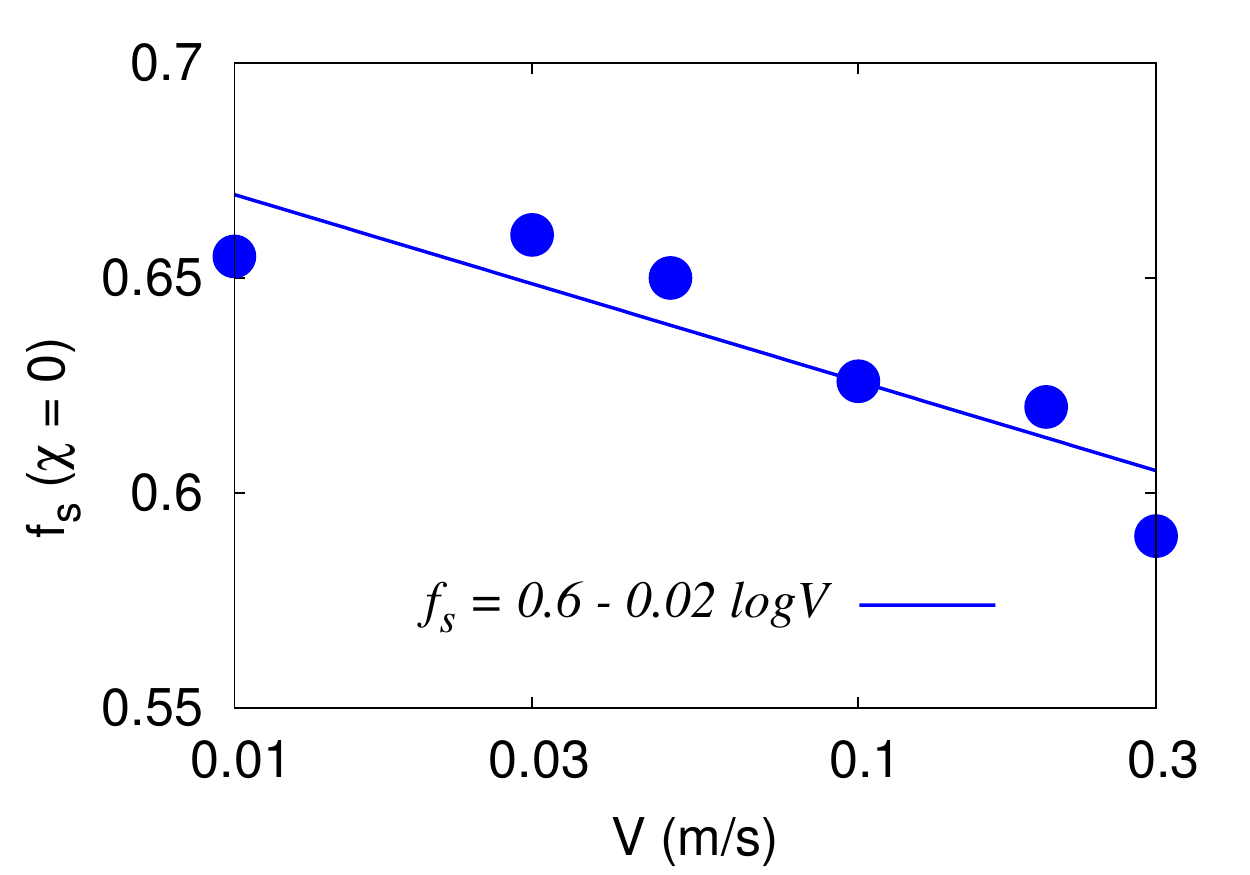}
\caption{Solid fraction $f_s$ is interpolated in the bridging plane in mushy zone where $\chi$ = 0. Cell roots and liquid channels coalesce when $f_s$ is between 0.6 and 0.7. The line of best fit for $f_s(\chi = 0)$ vs. $V$ is shown.}\label{fig_bridging}
\end{center}
\end{figure}

The characteristic temperature $T_{\mbox{\scriptsize bridge}}$ is extracted from the mushy zone in Fig.~\ref{fig_connection} and presented in Fig.~\ref{fig_tempb}, along with a reference to the liquidus and eutectic temperatures from the phase diagram~\cite{knorovsky1989}. It is evident that $T_{\mbox{\scriptsize bridge}}$ increases with decreasing V, which also signifies that more liquid is retained in the mushy zone with decreasing $V$. We extract the solid fraction $f_s$ corresponding to the $T_{\mbox{\scriptsize bridge}}$ isotherm, as shown in Fig.~\ref{fig_bridging}. The $f_s(T_{\mbox{\scriptsize bridge}})$ curves suggest that coalescence takes place when $f_s$ ranges from 0.6 to 0.7 in the mushy zone. When $f_s$ is greater than 0.7 in our simulations, there is a morphological change in the primary $\gamma$ phase from isolated solid clusters of the cells surrounded by liquid, to a continuous solid network with isolated liquid droplets. This type of transition from a solid-like to a liquid-like behavior in the mushy zone is very critical in the formation of solidification defects~\cite{Provatas2017,Zhou2005,Chen2016}. We have also noticed that over very short distances from the bridging plane in the mushy zone, $f_s$ increases rapidly to 0.9 and above. This is not surprising due to the rapid nature of AM solidification. It is worth noting that, when the temperature drops below $T_e$, any remaining liquid in the mushy zone becomes metastable and could transform to secondary eutectic phases. The residual liquid volume and temperature in the mushy zone may thus influence processes such as solid bridging in late solidification stages and precipitation of secondary phases in the solid-state. The present binary model however does not represent any phases beyond liquid and $\gamma$. Work in this direction is currently in progress using multi-phase-field approaches~\cite{supriyo2014,Ghosh2015,Ghosh2017_eutectic}.

\section{Discussion}\label{sec_discussion}
The present results can be used as a reference for simulation of AM microstructures using the quantitative phase-field model proposed in Refs.~\cite{Karma2001,Echebarria2004}. This model is based on significant simplifications of real AM experiments, such as the frozen-temperature approximation, absence of convection and absence of interface kinetics. Our results show that local equilibrium does not hold during AM solidification, although the anti-trapping solute flux term (in Eq.~(\ref{eq_c})) is meant to enforce local equilibrium. This contribution, in particular, needs refinement in order to quantify the interface-induced solute-trapping during AM processes. The fundamental solid-liquid interface properties such as surface energy and kinetic coefficients are also needed for more accurate simulations. 

We ignore the effects of convection on the primary arm spacing and Nb segregation. The related number for convective turbulent flow is the Prandtl number $\approx$ 0.3 and for thermal transport is the Lewis number $\approx$ 700 for Ni-Nb. One would expect these numbers to somewhat affect the microstructure evolution during rapid directional solidification, however the magnitude of these effects has not been tested presently in this study. Effects of convection on the primary arm spacing is not as pronounced compared to the sidebranches~\cite{Wang2003,Lee2010}, which are not observed in our simulations. Interestingly, the simulations performed by Lee \emph{et al.}~\cite{Lee2010} showed that the effects of convection are negligible where 3D simulations were concerned, and thus the solute partitioning across the interface remains similar when simulations were conducted with and without convection. Details of 2D and 3D simulations of dendritic solidification under convection are illustrated in Refs.~\cite{Lee2010,Eshraghi2017}. The effects of convection modify the Ivantsov function (Eqs.~(\ref{eq_iv3d}) and (\ref{eq_iv2d})) during rapid directional solidification, as discussed in Ref.~\cite{Galenko2017}. Moreover, diffusion of heat is ignored in the present work. This can be accommodated in a thermo-solutal phase-field model where a temperature field is also solved simultaneously along with $\phi$ and $c$~\cite{Mullis2010}. Approximation of temperature as a planar isotherm may be strongly idealized when compared to AM conditions.

The present work provides a qualitative basis for the connections and coalescence between cells belonging to the same grain. Simulations have been performed in bi-crystals, where the misorientation angles and the convergent/divergent growth conditions were considered between grains which have been found to affect the coalescence behavior in the mushy zone~\cite{Rappaz1999,Wang2004,Rappaz2003,Provatas2017}. In this regard, hot cracking models~\cite{Rappaz1999,Wang2004,Rappaz2003} coupled with phase-field simulations or phase-field crystal models~\cite{Wang2017} could capture the solidification shrinkage, mechanical strains, liquid cavitation, and microsegregation behavior in the mushy zone during late stage AM solidification in order to model microporosity, hot cracking and other deformation mechanisms within a dendritic framework. The solidification shrinkage stress was found to be larger with increasing velocity~\cite{Wang2017}, for which the coalescence begins at a smaller solid fraction. Late stage solidification behavior during equiaxed mode of solidification was found different than cellular mode of solidification in Ref.~\cite{Montiel2014}. 

Solute diffusion is very efficient in 3D and cell spacings are therefore smaller in 3D compared to 2D. The cell spacings simulated in the present work are smaller than \SI{0.5}{\micro\metre}. Such dense cellular microstructures provide significant resistance to fluid flow in experiments following an exponential increase of the damping effect in the mushy region and the effects of convection are therefore minimized~\cite{Yang2001,Tan2011}. The effect of fluid convection is less in the solidification or microstructure evolution front compared to vicinity of the molten pool surface in a laser melting process~\cite{Lee2014}. The number of primary cells remains same with and without fluid flow during the simulations reported in Refs.~\cite{Lee2010,Lee2014,Tan2011}. In addition, consideration of a dilute alloy reduces the effects of convection on the solute composition~\cite{Wang2003}. Neglecting convection in this work is therefore reasonable, given our focus on the average behavior of primary dendrites toward the selection of spacing, composition, temperature, and coalescence patterns. 

\section{Conclusions and perspectives}\label{sec_conclusions}
We have used a binary alloy phase-field model to simulate 3D AM cellular microstructures that form under the solidification conditions obtained from a (FEA simulated) 3D melt pool. A Ni-Nb binary alloy is modeled with properties approximating a Ni-based superalloy. Our simulation results show that cell spacings are finer in 3D in comparison to 2D and are approximated by $\lambda_c \propto G^{-m}V^{-n}$ with $m$ and $n$ between 0.25 and 0.5. Through the analysis of the composition fields, we show that microsegregation in 3D cell core is well approximated by the dendrite growth theory, while 2D simulation results deviate significantly from the theory. On the other hand, the simulated $T$ fields are close to the theory even in the AM solidification regime. Our results from the analysis of solidification pathways by $f_s$ vs. $T$ curves estimate the cumulative solid and liquid fractions in $x-y$ vertical sections. The Hoshen-Kopelman algorithm was then used to analyze the mushy zone from these sections in order to determine the characteristic bridging plane on the basis of zero Euler characteristic. The results from these analyses indicate that the onset of coalescence occurs when the solid fraction of $\gamma$ in the mushy zone is between 0.6 and 0.7. The residual liquid volume fraction in the mushy zone may have influence on the formation of solidification defects during late solidification stages. These results provide a basis in our ongoing efforts on predictive simulation of AM solidification microstructures.

The simulated microstructures and the concentration field can be used as inputs for the simulation of subsequent solid-state phase transformations. Secondary eutectic phases may arise in between the $\gamma$-cells because the Nb-rich droplets observed in the present simulations are expected to transform into Laves phase during the subsequent stages of solidification in IN718~\cite{nie2014,Kundin2015,Nastac1996}. Particularly for the discontinuous liquid droplets that emerge as a function of pinching off from the cell roots, an analysis of these structures can offer insight into the size and volume fraction of secondary phases~\cite{Ghosh2018}. We plan on molecular dynamics simulations to obtain realistic values of solid-liquid interfacial energy and kinetic coefficients for the phase-field simulations. Integrated modeling of solidification and solid-state transformations would help parameter-microstructure optimization and hence alloy development.

\section*{Acknowledgments}
We thank Li Ma for providing results from the finite element simulations and Greta Lindwall for the thermodynamic calculations. S.G. thanks William Boettinger, Kevin McReynolds and Eric Zhu for careful reading of the manuscript and constructive feedback. S.G. acknowledges Lyle Levine and Eric Lass for discussions of the experimental results at NIST that inspired the present work. N.O.-O. acknowledges the following financial assistance: Award No. 70NANB14H012 from U.S. Department of Commerce, National Institute of Standards and Technology as part of the Center for Hierarchical Materials Design (CHiMaD).
\section*{References}
\bibliography{papers}

\end{document}